# Pressure coefficients of Raman modes of carbon nanotubes resolved by chirality: Environmental effect on graphene sheet

A.J. Ghandour,[1,2] I.F. Crowe,[3] J.E. Proctor,[4] Y.W. Sun,[1] M.P. Halsall,[3] I. Hernandez,[1] A. Sapelkin[1] and D.J. Dunstan[1,*]

[1] School of Physics and Astronomy, Queen Mary University of London, London E1 4NS, UK.
[2] Department of Physics, Faculty of Sciences 5, Lebanese University, Nabatieh, Lebanon.
[3] Photon Sciences Institute and School of Electrical and Electronic Engineering, University of Manchester, Manchester M13 9PL, UK.
[4] Department of Physics, University of Hull, Hull HU6 7RX, UK.

Studies of the mechanical properties of single-walled carbon nanotubes are hindered by the availability only of ensembles of tubes with a range of diameters.  Tunable Raman excitation spectroscopy picks out identifiable tubes. Under high pressure, the radial breathing mode shows a strong environmental effect shown here to be largely independent of the nature of the environment . For the G-mode, the pressure coefficient varies with diameter consistent with the thick-wall tube model. However, results show an unexpectedly strong environmental effect on the pressure coefficients. Reappraisal of data for graphene and graphite gives the G-mode Grüneisen parameter $\gamma = 1.34$ and the shear deformation parameter $\beta = 1.34$.

Single-walled carbon nanotubes (SWCNTs) have great potential in applications ranging from nanofluidics to composite reinforcement. Consisting of a single, rolled-up graphene sheet, they have only surface atoms and so are unusually sensitive to their environment. This sensitivity hampers investigation of many intrinsic properties of the nanotubes; in particular, their response to high hydrostatic pressure.[1]

Raman spectroscopy has been used extensively for investigating the structural, mechanical and vibrational properties of SWCNTs. The Raman G-band, at about 1600cm$^{-1}$, derives from the bulk graphite in-plane E$_{2g}$ mode while the low-frequency radial breathing mode (RBM) is a consequence of the tube structure. The pressure dependence of these modes carries key information about the bond anharmonicity and the mechanical strength of the curved graphene sheet. However, the Raman signal is highly resonant and nanotube samples always contain a large number of different diameters and chiralities, denoted by the chiral indices (*n*, *m*). The Raman spectrum is dominated by those tubes whose electronic transition energies $E_{ii}$ match the laser excitation energy.[2] As well as shifting with pressure,[3, 4] the electronic transition energies are also highly sensitive to the nature of the solvent or hydrostatic pressure transmitting medium (PTM) in which the nanotubes are immersed.[3, 4, 5] The result is that different nanotubes are in resonance with any given laser excitation energy in different solvents, and with increasing pressure different tubes come in and out of resonance.[1] As a consequence, unambiguous determination of the pressure coefficients of the Raman peaks is complicated, and, most remarkably, no clear difference between (solvent) filled and empty tubes has yet been reported.[5]

A large body of published work has shown that resonant Raman spectroscopy of carbon nanotubes at ambient pressure, in which both the RBM shift $\omega_{RBM}$ and resonance energy $E_{ii}$ are measured, gives peaks on a two-dimensional surface to which chiral indices (*n*, *m*) can be assigned. This work began with the *Kataura* plot of theoretical $E_{ii}$ values against diameter for all (*n*, *m*).[2] More recent experimental and theoretical work refined this plot so that identification of many peaks from their ($\omega_{RBM}$, $E_{ii}$) position is now unambiguous.[6-10] Whilst the bulk of these studies concern unbundled nanotubes in water with surfactant, different shifts have been observed with different surfactants [6], and the effect of filling open tubes with water has also been reported.[10]

We have reported large shifts in the $E_{ii}$ co-ordinate of some (*n*, *m*) nanotubes in the form of bundles in different solvents (water, hexane, sulphuric acid) and in air.[3] In contrast, high pressure with water as the PTM (solvent) gives a shift which is largely in the $\omega_{RBM}$ coordinate.[3] This shows that solvent effects and pressure effects are distinct, and opens the way to obtaining reliable pressure coefficients for each (*n*, *m*), not only for the RBM mode but also for the G-mode. Here we demonstrate this by obtaining the RBM and G-mode pressure coefficients for three peaks in the ($\omega_{RBM}$, $E_{ii}$) map. Results for the RBM agree well with previous authors. For the G-mode the results are not as expected from the current interpretation of the pressure dependence of the graphene and graphite equivalents of the G-mode, an issue which we address here.

Hipco SWCNTs were used as bought, without unbundling, in water as the PTM. A Ti-sapphire laser was used to perform Raman spectroscopy over the energy range 1.48 eV – 1.78 eV at intervals of about 10meV. At each excitation energy, Raman spectra were recorded over the range 210 cm$^{-1}$ – 320 cm$^{-1}$ to capture the RBM peaks, and from 1500 cm$^{-1}$ – 1700 cm$^{-1}$ for the G-band spectra. The RBM spectra are fitted with Lorentzian peaks (giving the positions $\omega_{RBM}$), and the intensity of each peak is plotted against the laser wavelength. The laser excitation energy giving maximum RBM intensity is taken as $E_{ii}$ for that peak. We presented the *Kataura* plot thereby obtained, with the chiralities assigned by comparison with the results of Araujo *et al.*[8,9] in Ref. 3. Given the chiralities $(n, m)$, the diameters are calculated as $d = a\pi^{-1} (n^2 + nm + m^2)^{½}$ where the C-C bond length is $a = 0.246$ nm. At laser wavelengths near 1.75eV there is a single dominant peak in the RBM spectrum (Fig.1) that is assigned to the (9, 1) chirality ($d = 0.747$nm). At 1.64eV the peak assigned to the (11, 0) ($d = 0.861$nm) and (10, 2) ($d = 0.872$ nm) chiralities dominates the spectrum, and at laser wavelengths near 1.53eV it is the (12, 1) ($d = 0.981$nm) and (11, 3) ($d = 1.000$nm) peak which dominates. At most other excitation wavelengths there are two or more strong peaks in the RBM spectrum. When a single RBM peak dominates the spectrum, it is likely that the G-band peak will be largely due to the same chirality or chiralities, while if there are two or more strong peaks in the RBM spectrum then the G-band peak would contain contributions from each. Consequently, we focus our attention here on these three excitation energies.

The pressure experiments were carried out in a diamond-anvil cell operated in the Zen configuration (using a single diamond[11]), which permits good control over the pressure in the range 0 - 2GPa. The pressure was measured using the standard technique of ruby photoluminescence.

The dependence of the RBM spectra and the G-band spectra on pressure is shown in Fig.1 for the three excitation wavelengths. The peak positions are plotted in Fig.2 with linear least-squares fits to obtain the pressure coefficients. To estimate the errors due to scatter, and also because the 2GPa points may have increased error due to the freezing of the water PTM above 1GPa, least-squares fits to the data for the three lower pressures are also shown.

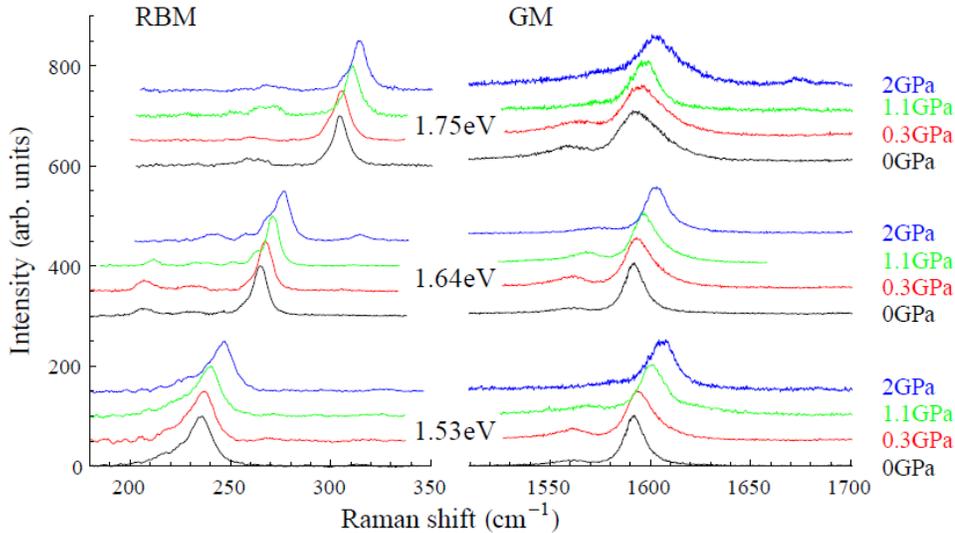

FIG. 1. RBM and G-mode spectra for the excitation energies and pressures marked, offset vertically for clarity. The spectra under 1.75eV excitation (upper group) are assigned to the (9, 1) chirality, the spectra under 1.64eV excitation (middle group) to the (11, 0) and (10, 2) chiralities and the spectra under 1.53eV (lower group) to the (12, 1) and (11, 3) chiralities.

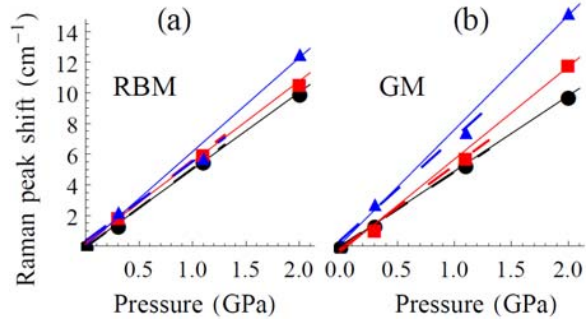

FIG. 2. Shifts with pressure for (a) the RBM peaks, and (b) the G-mode peaks, for the three excitation energies of Fig.1. The solid lines are linear least-squares fits to the whole datasets, while the dashed lines are fits to the lower three pressure points.

In Fig. 3, the pressure coefficients we measure are plotted against the tube diameters, using solid circles for the fits to the lower pressure points and open circles for the fits that include the 2GPa data. Literature data for the pressure coefficients of the RBM peaks of semiconducting SWCNTs is also shown. Experimental data is for the RBM of bundled tubes in an ethanol-methanol mixture from Venkataswaran et al.[12] and the RBM of unbundled tubes in $H_2O$ with surfactant from Lebedkin et al.[13] Simulation data is for molecular dynamics (MD) of the RBM of isolated tubes in $H_2O$ from Longhurst and Quirke.[14]

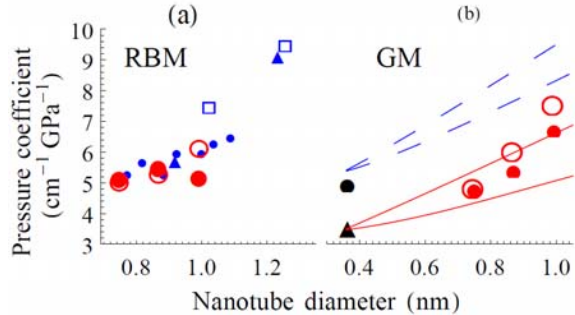

FIG. 3. (Colour online.) Pressure coefficients from Fig.2 plotted against the tube diameters for the three excitation energies of Fig.1. In (a), the RBM data (large circles) are compared with experimental results for bundled semiconducting tubes in ethanol/methanol (triangles),[12] unbundled semiconducting tubes in water/surfactant (small solid circles),[13] and with the MD simulation results for unbundled semiconducting tubes in water from Ref.14 (open squares). In (b), the G-mode data are plotted. The broken lines show the dependence on diameter expected for the $G^+$ and $G^-$ bands from Eqn.3 with the values for $\gamma$ and $\beta$ given in Ref.18 and the solid lines show the results for the revised values discussed in the text. For comparison, the pressure coefficients of graphite (solid circle)[19] and graphene (solid triangle) (revised value from the data of Ref.18 according to Eq.2 with $\varepsilon_T = 0$) are shown, plotted at $d = w$.

A striking feature of the results in Fig. 3(a) is the excellent agreement of our RBM data with the data for semiconducting debundled tubes of Lebedkin et al.[13] and for the bundled tubes of Venkateswaran et al.[12] Previously, differences in reported pressure coefficients were attributed to consequences of bundling (e.g. hexagonalisation under

pressure[12]) and to the different solvents used as PTM.[1] The good agreement between bundled and unbundled tubes in water and unbundled tubes in ethanol-methanol suggests that neither of these factors affects the pressure coefficients. This is a surprising but useful result.

The RBM frequency has been related to the G-band frequency by Venkateswaran *et al.*[12] using a continuous elastic medium approximation and by Gerber *et al.*[15] using a simple ball-and-spring model. In both analyses good agreement is obtained with the empirical dependence of $\omega_{RBM}$ on diameter. Both analyses imply a small RBM pressure coefficient of about $0.8d^{-1}$ cm$^{-1}$GPa$^{-1}$ where the tube diameter $d$ is in nm (before correction for the thick-wall effect, see Eq.1 below), very much less than the values observed. The MD simulations of Longhurst and Quirke[14] explain this in terms of the interaction between the (unbundled) nanotube and its environment by considering a nanotube surrounded by water molecules at high pressure. The van der Waals interaction between the nanotube and the first shell of water molecules provides only a small correction to the ambient-pressure RBM frequency, but the increase in the force constant of this interaction with pressure gives the bulk of the RBM pressure coefficient. This is a greater effect for low RBM frequencies (large tubes) than for high (small tubes), giving the dependence of the pressure coefficient on the diameter seen in Fig. 3(a). The good agreement of the data for bundled tubes in water, unbundled tubes in water and surfactant, and bundled tubes in ethanol-methanol suggests that the increase in the force constant of the interaction between the nanotube and its environment is similar in all cases. It would seem that the same RBM pressure coefficient (within experimental error) is obtained by the stiffening of the inter-nanotube van der Waals interaction in nanotube bundles as by the stiffening of the water (or surfactant) van der Waals interaction with unbundled tubes.

The G-band pressure coefficients in Fig.3(b) are remarkably low – in this low-pressure range, values up to 8 or 10 cm$^{-1}$GPa$^{-1}$ have commonly been reported[1] – and vary quite fast with diameter. The dependence on diameter may be understood by considering the nanotube as a thick-walled closed tube under external pressure $P$.[16,17] For an outside diameter of $d + w$ and an inside diameter of $d - w$, with $d > w$, the axial and tangential stresses are greater than the pressure $P$,

$$\sigma_L = \frac{(d+w)^2}{4dw}P, \quad \sigma_T = \frac{d+w}{2w}P \qquad (1)$$

These are unequal, so to predict the pressure coefficient we require both the hydrostatic and the shear deformation parameters (mode Grüneisen parameters) $\gamma$ and $\beta$. These are available from the experimental data of Mohiuddin *et al.*,[18] who studied the Raman G-band in graphene as a function of uniaxial strain, obtained by flexure of a beam to which a graphene flake adhered. Under uniaxial strain, the G-band splits into two components, $G^+$ and $G^-$. Dropping unnecessary notation and combining their Eq.3 with their experimental results, they gave

$$\omega^{G^\pm}_{\varepsilon_L} = \frac{\partial \omega^{G^\pm}}{\partial \varepsilon_L} = -2125 \mp 1045 \text{cm}^{-1} = -\omega_0^G \gamma(\varepsilon_L + \varepsilon_T) \pm \tfrac{1}{2}\omega_0^G \beta(\varepsilon_L - \varepsilon_T) \qquad (2)$$

where $\varepsilon_L$ is the longitudinal strain imposed on the graphene flake by the curvature of the substrate beam. They used the Poisson ratio $\nu = 0.33$ of the substrate to obtain the transverse strain $\varepsilon_T = -\nu\varepsilon_L$, and, using the experimental value of $\omega_0 = 1590$cm$^{-1}$ for the G-band frequency at ambient pressure, they obtained the G-mode parameters as $\gamma = 1.99$ and $\beta = 0.99$. The hydrostatic strain coefficient of graphene under hydrostatic pressure $P$ is $\omega_\varepsilon^{G\pm} = 2\omega_0\gamma = -6340 \text{cm}^{-1}$, which, with $(s_{11} + s_{12})^{-1} = 1250 \text{GPa}$, corresponds to $\omega_P^{G\pm} = 5.07 \text{ cm}^{-1}\text{GPa}^{-1}$ in good agreement with experimental values for graphite [19]. However, using $\varepsilon_T = -\nu\varepsilon_L$ for the transverse strain is incorrect. For a thin beam in flexure, as the tensile part above the neutral plane tries to contract laterally and the compressive part below tries to expand, there is no transverse strain – i.e. this is a plane strain problem.[20] To correct this, put $\varepsilon_T = 0$ in Eq.2, giving $\gamma = 1.34$ and $\beta = 1.31$, or to experimental accuracy, $\gamma \sim \beta \sim 4/3$. Then the predicted pressure coefficient for graphene and graphite from the experimental data of Ref.18 becomes $3.40 \text{ cm}^{-1}\text{GPa}^{-1}$.

For the nanotube, using Eq.1 for the axial and tangential stresses under a pressure $P$ and taking $\nu = 0.13 = -s_{12}/s_{11}$, $s_{11} + s_{12} = 1/1250$ GPa as in Ref.18, the strains and the pressure coefficients of the $G^\pm$ bands are given by

$$\varepsilon_L = s_{11}\sigma_L + s_{12}\sigma_T, \quad \varepsilon_T = s_{12}\sigma_L + s_{11}\sigma_T$$
$$\varepsilon_H = \varepsilon_L + \varepsilon_T, \quad \varepsilon_S = \varepsilon_L - \varepsilon_T \qquad (3)$$
$$\omega_P^{G\pm} = \omega_0\gamma\varepsilon_H \mp \tfrac{1}{2}\omega_0\beta\varepsilon_S$$

These curves are plotted in Fig.3(b) against $d$ for $w = 0.36$nm for the values of $\gamma = 1.99$ and $\beta = 0.99$[18] (broken curves) and these do not agree with the data. They are plotted also for the revised values of $\gamma = 1.34$ and $\beta = 1.31$ (solid curves) and these show good agreement with the data, within experimental uncertainty.

These results are surprising. With this revision of the result of the uniaxial experiment of Mohiudden et al.[18] we have good agreement between their data and the data for nanotubes under high pressure. On the other hand, these results are in sharp disagreement with data for graphene and graphite under high pressure, where much higher pressure coefficients are reported. Initial experiments on graphene under hydrostatic pressure[21] gave G-mode peak shifts as a function of strain / pressure that were consistent with density-functional (DFT) calculations[21] and simple mechanical models assuming that the Raman peak shifts are due entirely to the bond stiffening when the C-C decreases. However, more recent experimental results[22] showed the graphene G-mode pressure coefficient varying from 8-11$\text{cm}^{-1}\text{GPa}^{-1}$ according to the choice of PTM, as observed in nanotubes.

If a significant part of the G-mode pressure coefficient derives from interaction with the environment, then it is noteworthy that the uniaxial stress experiment on graphene and nanotubes under pressure (whether bundled or unbundled) have condensed matter (solid or liquid) in contact with one side only of the graphene sheet. In contrast, graphene under pressure and graphite both have condensed matter (solid or liquid) in contact with both sides of each graphene sheet. Without speculating on the origin of the environmental effect, there is scope for it being twice as large in this case. This requires that a significant part of the graphite pressure coefficient is due to interactions between the graphene sheets (each sheet serving as part of the environment of its neighbours).

This interpretation also predicts that open tubes which fill with PTM will display a higher pressure coefficient than expected from the data for closed tubes (but independent of diameter). This may explain why no clear difference has been reported between closed tubes, with pressure coefficients raised by the thick-wall effect (Eq.1), and open tubes.[5]

In pressure experiments on double-walled nanotubes-23, 24, the inner tube has condensed matter on one side only, while the outer tube has it on both sides. The pressure coefficients of the inner tubes (3.3-5.1cm$^{-1}$GPa$^{-1}$) are consistently much lower that those of the outer tubes (5.8-8.6cm$^{-1}$GPa$^{-1}$).[24] The data were interpreted in terms of the intertube pressure,[23] but the data are also consistent with the environmental effect suggested here.

The data reported here utilise tunable laser excitation to obtain the first reliable pressure coefficients for both the Raman modes of individual single-walled carbon nanotubes that may be assigned to chirality and diameter. Experimentally, it is clearly urgent to find the G-mode pressure coefficients for nanotubes for a larger range of diameters, in different solvents, and for open tubes as well as closed. The results for the RBM show that the increase in the force constant of the interaction between the nanotube and its immediate surroundings at high pressure occurs in a similar manner for tubes surrounded by other nanotubes, surfactant or solvent. The results for the G-band are unexpected and have stimulated a correction of the available data for graphene. Theoretically, they suggest the calculation *ab initio* of graphene, when the π-orbitals are compressed by an adjacent graphite layer or PTM on one side and on both sides. This study represents a major step forward to achieving a unified understanding of the characteristics of graphene-based structures under stress and gives clear guidance as to what further studies are necessary to complete this understanding.

**Acknowledgements:** We acknowledge financial support from the Engineering and Physical Sciences Research Council.